\documentclass[prb,twocolumn,showpacs,amsmath,amssymb,superscriptaddress]{revtex4}
\usepackage[dvips]{graphicx}% Include figure files--uses eps figures
\usepackage{graphics}% Include figure files
\usepackage{dcolumn}% Align table columns on decimal point
\usepackage{bm}% bold math

\newcommand{\znpk}{z_{n'k}}
\newcommand{\znk}{z_{nk}}
\newcommand{\kk}{{\bf k}}

\begin{document}
\title{Infrared magneto-optical properties of (III,Mn)V ferromagetic semiconductors}  
%\title{Anomalous Hall effect and large magneto-optical effects at
%infra-red frequencies in ferromagnetic (III,Mn)V semiconductors}
\author{Jairo Sinova}
\affiliation{University  of  Texas at  Austin,  Physics Department,  1
University Station C1600,  Austin TX 78712-0264} 
\author{T. Jungwirth}
\affiliation{University  of  Texas at  Austin,  Physics Department,  1
University   Station  C1600,   Austin  TX   78712-0264}  
\affiliation{ Institute of Physics  ASCR, Cukrovarnick\'a 10, 162 53  
Praha 6, Czech Republic } 
\author{J. Ku\v{c}era}
\affiliation{ Institute of Physics  ASCR, Cukrovarnick\'a 10, 162 53
Praha 6, Czech Republic }
\author{A.H. MacDonald} \affiliation{University of Texas at
Austin,  Physics Department,  1  University Station  C1600, Austin  TX
78712-0264} 
\date{\today}
\begin{abstract}
We present a theoretical study of the infrared magneto-optical properties of
ferromagnetic (III,Mn)V semiconductors.
Our analysis combines the kinetic exchange model for (III,Mn)V ferromagnetism
with Kubo linear response theory and Born approximation estimates for 
the effect of disorder on the valence band quasiparticles.
We predict a prominent feature in the ac-Hall conductivity at a 
frequency that varies over the range from 200 to 400 meV, depending on 
Mn and carrier densities, and is associated with transitions between heavy-hole
and light-hole bands.  In its zero frequency limit, our Hall conductivity reduces
to the $\vec k$-space Berry's phase value predicted by a recent theory of the 
anomalous Hall effect that is able to account quantitatively for experiment.
We compute theoretical estimates for magnetic circular dichroism, Faraday rotation, and
Kerr effect parameters as a function of Mn concentration and free carrier density.
The mid-infrared response feature is present in each of these magneto-optical 
effects.

\end{abstract}

\maketitle %it has to be after abstract and pacs

\section{Introduction}

Rapid progress has  been achieved over the past  year in understanding
how  growth  and  annealing  conditions influence  the  properties  of
(III,Mn)V diluted magnetic semiconductor ferromagnets.  These advances
have  led to  the  realization of  samples  with higher  ferromagnetic
transition          temperatures          and          conductivities.
\cite{edmonds_apl02,ku_cm0210426,kuryliszyn_cm0207354}        (III,Mn)V
materials  have  normally  been  described  using  a  phenomenological
model\cite{bookchapter,  dietl_prb01,dietl_science00}   in  which  the
valence band  holes of the  host (III,V) semiconductor are  coupled by
exchange and Coulomb interactions  to Mn$^{2+}$ local-moment ions with
spin $S=5/2$.  The properties predicted by this model are most simply
understood in the  strongly metallic regime for which  disorder in the
spatial distribution of  the Mn$^{2+}$ ions, and other  defects of the
materials, can  be treated perturbatively.   These approximations lead
to  a picture of  the materials  in which  spin-orbit coupling  of the
valence-band  hole subsystem  plays a  key role  \cite{bookchapter} in
providing   detailed  explanations   for   many  qualitative   effects
discovered  in  experimental studies  of  thermodynamic and  transport
phenomena.   The model  can  account quantitatively  for the  critical
temperature,   \cite{dietl_prb01,jungwirth_prb02}   strain   sensitive
magnetic-crystalline   anisotropy,   \cite{abolfath_prb01,dietl_prb01}
anisotropic   magneto-resistance  coefficients\cite{jungwirth_apl02},
and the  strong anomalous Hall effect.\cite{jungwirth_prl02,sinova_cm}
Golden  rule  estimates of  quasiparticle  scattering amplitudes  even
provide   the  correct   order  of   magnitude  for   longitudinal  dc
conductivities.\cite{jungwirth_apl02}   

In  this   paper   we  discuss
corresponding theoretical predictions for the infrared magneto-optical
properties  of   these  materials.   We   evaluate  magnetic  circular
dichroism (MCD),  Faraday rotation, and  Kerr effects in  the infrared
regime for several different  Mn concentrations and carrier densities.
From the microscopic point of view, each of these effects reflects the
non-zero  value of  the ac  Hall  conductivity, $\sigma_{xy}(\omega)$.
Our linear  response theory for  the Hall conductivity reduces  in the
zero frequency  limit to the  $\vec k$-space Berry's  phase expression
that       explains      dc       Hall       effect      observations.
\cite{jungwirth_prl02,sinova_cm,sundaram_prb99}

In metallic ferromagnets, measurements of magneto-optical coefficients
on  band energy  scales provide  very detailed  information  about the
influence  of  broken  time-reversal  symmetry on  itinerant  electron
quasiparticle  states.   The appropriate  band  energy  scale for  the
heavily  p-doped (III,Mn)V  ferromagnets, and  for a  number  of other
materials  that have  been studied  recently\cite{newrefs}, is  in the
infrared.   For this  reason,  we believe  that experimental  infrared
magneto-optical   studies  of   (III,Mn)V   ferromagnets  are   highly
desirable; we expect that they will  be carried out in the near future
and that comparison  with the predictions presented here  will be very
informative  in  clarifying  the  physics  of  these    new
ferromagnets.   They could,  for example,  reveal deficiencies  of the
relatively simple  theoretical formulation that we  employ.  The study
of  the  magneto-optical  response   of  these  ferromagnets  is  also
potentially   interesting  for   applications,   especially  if   room
temperature ferromagnetism is achieved in the future.  Magneto-optical
properties   of  the  closely   related  (II,Mn)VI   diluted  magnetic
semiconductor paramagnets \cite{ando_book}  have already proved useful
from both basic science and application points of view.

Absorption and reflection measurements  in the visible range have been
used  to establish  phenomenological  estimates for  the  p-d and  s-d
exchange   coupling   constants  in   (II,Mn)VI   materials,  and   in
establishing  the  important  role   of  valence  band  holes  in  the
(III,Mn)V's.
\cite{dietl_prb01,ando_book,ando_jap98,szczytko_prb99,beschoten_prl99}
Photoemission  experiments,   which  explore  the   deeper  electronic
structure  have  been used  to  explore  the  degree of  hybridization
between the underlying host valence band and Mn electronic levels, but
suffer          from         being          surface         sensitive.
\cite{okobayashi_prb99,okobayashi_prb02,okobayashi_prb01}    In    the
infrared  regime,   recent  optical  conductivity   measurements  have
uncovered unusual non-drude  behavior, including an optical absorption
peak   \cite{singley_prl02,hirakawa_prb02,katsumoto_mseb01,nagai_jjp01}
connected to back-scattering localization effects and to inter valence
band    transitions,   in    agreement   with    model   calculations.
\cite{sinova_prb02,yang_cm0202021}

We organize  the paper as  follows. In Sec.~\ref{TA} we  introduce the
Kubo  formula  description  of  the  ac  anomalous  Hall  conductivity
appropriate for the studied  (III,Mn)V ferromagnets.  In Sec. \ref{KL}
we  detail  the  model  Hamiltonian  and approximations  used  in  our
calculations.  In  Sec. \ref{4band} we present  an analytic evaluation
of the anomalous  Hall conductivity for the case  in which disorder is
neglected and  the bands are  approximated by the  four-band spherical
model.   (The six-band  model that  we use  for  numerical calculation
reduces  to the four-band model  in  the limit  of infinite  spin-orbit
coupling strength.)  The isotropic band dispersion of this model makes
analytic  calculations  possible,  although  they are  still  somewhat
cumbersome.  The  details of this calculation,  which builds intuition
about qualitative properties  of the $\sigma_{xy}(\omega)$ curves, are
relegated  to  an  appendix.   In  Sec.  \ref{6band}  we  present  the
numerical  results  of  the  full model  Hamiltonian  calculation  for
$\sigma_{xy}(\omega)$  and  apply these  results  to  discuss all  the
common  magneto-optical  effects  available  for  experiments  in  the
present geometry.  We summarize  this work and present our conclusions
in Sec. \ref{conclusion}.

\section{Theoretical approach}
\label{TA}
Our  theoretical  model  description   starts  by  coupling  the  host
semiconductor   valence   band   electrons,   described   within   the
$\vec{k}\cdot\vec{p}$ or  Kohn-Luttinger (KL) theory,  with $S=5/2$ Mn
local moments with  a semi-phenomenological local exchange interaction
treated  at   a  mean-field  level.  \cite{dietl_prb01,abolfath_prb01,
jungwirth_apl02,sinova_prb02} At  zero temperature this  gives rise to
valence  bands   that  are  split  by  an   effective  exchange  field
$\vec{h}=N_{Mn^{2+}}S  J_{pd}\hat{z}$,   where  $N_{Mn^{2+}}$  is  the
substitutional Mn density and the strength of the exchange coupling is
taken to  be $J_{pd}=55$ meV nm$^{-3}$.\cite{ohno_jmmm99}  
We assume in  this paper that
the magnetization is aligned along the growth ($\hat{z}$) direction by
an  applied small external  magnetic field.   We restrict  ourselves  to the
$T=0$   limit,  allowing   us  to   neglect  scattering   off  thermal
fluctuations in the Mn moments orientation.  We assume collinear
magnetization  in  the  ground  state,  ignoring  the  possibility  of
disorder induced  non-collinearity in the ground state  which is known
to be less likely for  the strongly metallic (III,Mn)V ferromagnets on
which we focus.  \cite{schliemann_prl02,zarand_prl02}

% Comment: Caveats stated more explicitly here.  OK? 
% The operators p_x etc. were not defined previously.  OK to just cite other paper?  

The linear response theory Kubo formula expression for the real
part of the ac Hall conductivity of disorder-free non-interacting electrons is:
\begin{eqnarray}
{\rm Re}[\sigma_{xy}(\omega)]
&=&-\frac{e^2\hbar}{m^2 }\int \frac{d\vec{k}}{(2\pi)^3}
\sum_{n\ne n'} (f_{n',\vec{k}}-f_{n,\vec{k}}) \nonumber\\&&\times
\frac{{\rm Im}[\langle n' \vec{k}|\hat{p}_x|n\vec{k}\rangle\langle
n\vec{k}| \hat{p}_y|n' \vec{k}\rangle]}
{(\omega-E_{n\vec{k}}+E_{n'\vec{k}})(E_{n\vec{k}}-E_{n'\vec{k}})},
\label{ac_sig_AH}
\end{eqnarray}
where  $|n\vec{k}\rangle$  are   the  Bloch  valence-band  states  and
$E_{n\vec{k}}$  the Bloch  eigenenergies  within $\vec{k}\cdot\vec{p}$
theory (we use  either six or four band models here),  $m$ is the bare
electron mass, $f_{n,\vec{k}}$ is the  Fermi occupation number (0 or 1
at  $T=0$) for the  state $|n\vec{k}\rangle$,  and $\hat{p}/m$  is the
$\vec{k}\cdot\vec{p}$  velocity operator obtained \cite{yang_cm0210149}
by differentiating the  $\vec{k}\cdot\vec{p}$ Hamiltonian with respect
to  wavevector.   In  the  zero  frequency  limit  Eq.~\ref{ac_sig_AH}
reduces   to  the   expression  used   by  Jungwirth   {\it   et  al}.
\cite{jungwirth_prl02,sinova_cm}  to  explain  the dc  anomalous  Hall
conductivity  of  these materials.   This  recent  work suggests  that
anomalous   Hall   effects   are   more   quantitatively   useful   in
characterizing  itinerant electron  ferromagnets  than had  previously
been thought,  at least for the  present materials.  In  this paper we
extend this advance to finite frequencies.

Even  though  the  Hall  conductivity  is finite  in  the  absence  of
disorder,  we   do  anticipate   that  disorder  will   influence  the
$\sigma^{\rm  AH}_{xy}(\omega)$ curves,  primarily  by broadening  out
features.   The sources  of disorder  known  to be  relevant in  these
materials include positional randomness  of the substitutional Mn ions
with charge $Q=-e$, random placement of interstitial Mn ions which act
as      double     donors     and      are     believed      to     be
non-participants\cite{blinowski_cm0212093} in the ferromagnetic order,
and  As anti-sites which  also act  as having  charge $Q=+2e$  and are
non-magnetic.  We  estimate the influence  of disorder on  the valence
band  quasiparticles  by  calculating  their lifetimes  using  Fermi's
golden rule including both  screened Coulomb and exchange interactions
of  the  valence electrons  with  the  Mn  ions and  the  compensating
defects. \cite{jungwirth_apl02}  Including disorder broadening  of the
quasiparticle spectral functions, the  Kubo formula expression for the
Hall conductivity becomes
\begin{widetext}
\begin{eqnarray}
{\rm Re}[\sigma_{xy}(\omega)]=-\frac{e^2\hbar}{\omega m^2 V}
\sum_{\vec{k}n\ne n'}
{\rm Im}[\langle n'\vec{k}|\hat{p}_x|n\vec{k}\rangle
\langle n\vec{k}| \hat{p}_y|n'\vec{k}\rangle]\nonumber\\\times\int \frac{d\epsilon}{2\pi}
f(\epsilon) A_{n',\vec{k}}(\epsilon) {\rm Re}
[G^{\rm ret}_{n,k}(\epsilon+\hbar\omega)+f(\epsilon)A_{n,\vec{k}}(\epsilon)
{\rm Re}[G^{\rm adv}_{n',k}(\epsilon-\hbar\omega)],
\label{dis_sigma}
\end{eqnarray}
\end{widetext}
where                                         $A_{n,\vec{k}}(\epsilon)=
\Gamma_{n\vec{k}}/((\epsilon-E_{n\vec{k}})^2+\Gamma_{n\vec{k}}^2/4)$
is  the disorder  broadened spectral  function and  $G^{\rm  ret}$ and
$G^{\rm  adv}$ are  the  advanced and  retarded quasiparticle  Green's
functions  with finite  lifetime  $\Gamma_{n\vec{k}}^{-1}/2$, obtained
from the golden rule  scattering rates from uncorrelated disorder (see
Sec.  II). Since  we  are interested  in  the first  order effects  of
disorder in $\sigma_{xy}$ we approximate the above expression as,
\begin{widetext}
\begin{eqnarray}
{\rm Re}[\sigma_{xy}(\omega)]=-\frac{e^2\hbar}{m^2 V }% \int \frac{d\vec{k}}{(2\pi)^3}
\sum_{\vec{k}n\ne n'}
(f_{n',\vec{k}}-f_{n,\vec{k}}) 
%\nonumber\\ &\times& 
\frac{{\rm Im}[\langle n'
\vec{k}|\hat{p}_x|n\vec{k}\rangle\langle n\vec{k}| \hat{p}_y|n'\vec{k}
\rangle]
%\nonumber\\ &\times&
%\frac{
(\Gamma_{n,n'}^2+\omega(E_{n\vec{k}}-E_{n,\vec{k}'})
-(E_{n\vec{k}}-E_{n,\vec{k}'})^2)}
{((\omega-E_{n\vec{k}}+E_{n'\vec{k}})^2+\Gamma_{n,n'}^2)
((E_{n\vec{k}}-E_{n'\vec{k}})^2+\Gamma_{n,n'}^2)},
\label{ac_sig_AHE_dis}
\end{eqnarray}
\end{widetext}
where   $\Gamma_{n,n'}\equiv   (\Gamma_{n}+   \Gamma_{n'})/2   $   and
$\Gamma_n$ are the golden rule scattering rates averaged over band $n$
as      in      Ref.      \onlinecite{sinova_prb02}.       We      use
Eq.~(\ref{ac_sig_AHE_dis}) to evaluate $\sigma_{xy}(\omega)$ below.

\section{model hamiltonian}
\label{KL}
In the virtual crystal approximation, the interactions are replaced by
their spatial  averages, so that the Coulomb  interaction vanishes and
hole    quasiparticles   interact    with    a   spatially    constant
kinetic-exchange  field.  The  unperturbed Hamiltonian  for  the holes
then reads $H_0=H^L+\vec{h} \cdot\vec s\;$, where
$H_h$   is  the   host  band   Hamiltonian,  and   $\vec  s$   is  the
envelope-function  hole spin  operator.   The host  band  part of  the
Hamiltonian  is described  via  the four  or  six band  Kohn-Luttinger
model.   Choosing the  angular momentum  quantization direction  to be
along  the $  z$-axis,  and  ordering the  $j=3/2$  and $j=1/2$  basis
functions according  to the list  ($-3/2,1/2,-1/2,-3/2;1/2,-1/2$), the
Luttinger Hamiltonian $\hat{H}^L$ has the form \cite{abolfath_prb01}:
\begin{equation}
\hat{H}^L = \left(\begin{array}{cccccc} {\cal H}_{hh} & -c & -b &
\multicolumn{1}{c|}{0} & \frac{b}{\sqrt{2}} & c\sqrt{2}\\ -c^* & {\cal H}
_{lh}
& 0 & \multicolumn{1}{c|}{b} & -\frac{b^*\sqrt{3}}{\sqrt{2}} & -d\\ -b^*
& 0 &
{\cal H}_{lh} & \multicolumn{1}{c|}{-c} &   d & -\frac{b\sqrt{3}}{\sqrt{2
}} \\
0 & b^* & -c^* & \multicolumn{1}{c|}{{\cal H}_{hh}} &  -c^*\sqrt{2} &
\frac{b^*}{\sqrt{2}}\\ \cline{1-4} \frac{b^*}{\sqrt{2}} &
-\frac{b\sqrt{3}}{\sqrt{2}} & d^* & -c\sqrt{2} & {\cal H}_{so} & 0\\
c^*\sqrt{2} & -d^* & -\frac{b^*\sqrt{3}}{\sqrt{2}} & \frac{b}{\sqrt{2}} &
 0 &
{\cal H}_{so}\\
\end{array}\right)
\label{hl}
\end{equation}
In the matrix (\ref{hl}) we have highlighted the $j=3/2$ sector.
The Kohn-Luttinger eigenenergies are measured down from the top of the valence
band, i.e. they are hole energies.  For completeness we list the expressions 
which define the quantities that appear in $\hat{H}^L$:
\begin{eqnarray}
{\cal H}_{hh} &=& \frac{\hbar^2}{2m}\big[(\gamma_1 + \gamma_2)(k_x^2+k_y^
2) +
(\gamma_1 - 2\gamma_2)k_z^2 \nonumber,
\end{eqnarray}
\begin{eqnarray}
{\cal H}_{lh} &=&
\frac{\hbar^2}{2m}\big[(\gamma_1 - \gamma_2)(k_x^2+k_y^2) + (\gamma_1 +
2\gamma_2)k_z^2 \nonumber, \\ {\cal H}_{so} &=&
\frac{\hbar^2}{2m}\gamma_1(k_x^2+k_y^2+k_z^2) + \Delta_{so} \nonumber,
\end{eqnarray}
\begin{eqnarray}
b &=&
\frac{\sqrt{3}\hbar^2}{m} \gamma_3 k_z (k_x - i k_y) \nonumber, \\ c &=&
\frac{\sqrt{3}\hbar^2}{2m}\big[\gamma_2(k_x^2 - k_y^2) - 2i\gamma_3 k_x
k_y\big] \nonumber,
\end{eqnarray}
\begin{eqnarray}
d &=&
-\frac{\sqrt{2}\hbar^2}{2m}\gamma_2\big[2k_z^2-(k_x^2 + k_y^2 )\big]\; .
\label{lutpar}
\end{eqnarray}
We focus here on GaAs for which $\gamma_1=6.98$, $\gamma_2=2.06$, 
$\gamma_3=2.93$, and $\Delta_{so}=341$ meV . \cite{note2}

We treat the effects of disorder on the hole quasiparticles through
a finite lifetime scattering rate $\Gamma_{n\vec{k}}$ calculated using
the Fermi's golden rule. For uncorrelated disorder there are two 
contributions to the transport weighted scattering
rate $\Gamma_{n\vec{k}}=
\Gamma_{n\vec{k}}^{Mn^{2+}}+\Gamma_{n\vec{k}}^{As-anti}$ due
to substitutional Mn impurities and As-antisites, given by
\begin{eqnarray}
\Gamma^{Mn^{2+}}_{n,\vec k}&=&\frac{2\pi}{\hbar} N_{Mn^{2+}}\sum_{n^{\prime}}
\int\frac{d\vec k^{\prime}}{(2\pi)^3}
|M_{n,n^{\prime}}^{\vec k,\vec k ^{\prime}}|^2
\nonumber \\ &\times &\delta(E_{n,\vec k}
-E_{n^{\prime}\vec k ^{\prime}})
(1-\cos \theta_{\vec k, \vec k ^{\prime}})\nonumber\; ,
\label{gamma}
\end{eqnarray}
and
\begin{eqnarray}
\Gamma^{As-anti}_{n,\vec k}&=&\frac{2\pi}{\hbar} N_{As-anti}\sum_{n^{\prime}}
\int\frac{d\vec k^{\prime}}{(2\pi)^3}
|\tilde{M}_{n,n^{\prime}}^{\vec k,\vec k ^{\prime}}|^2
\nonumber \\ &\times &\delta(E_{n,\vec k}
-E_{n^{\prime}\vec k ^{\prime}})
(1-\cos \theta_{\vec k, \vec k ^{\prime}})\nonumber\; ,
\label{gamma_anti}
\end{eqnarray}
where the scattering matrix elements are approximated 
by the expressions (in S.I. units),
\begin{eqnarray}
M_{n,n^{\prime}}^{\vec k,\vec k ^{\prime}}&=&
J_{pd}S
\langle z_{n \vec k}|\hat{s}_z|
z_{n^{\prime}\vec k ^{\prime}}\rangle\nonumber \\
&-&
\frac{e^2}{\epsilon_{host}\epsilon_0(|\vec k -\vec k ^{\prime}|^2
+q_{TF}^2)}\langle z_{n \vec k}|
z_{n^{\prime}\vec k ^{\prime}}\rangle\nonumber, 
\label{mnelement}
\end{eqnarray}
and
\begin{eqnarray}
\tilde{M}_{n,n^{\prime}}^{\vec k,\vec k ^{\prime}}&=&
\frac{e^2}{\epsilon_{host}\epsilon_0(|\vec k -\vec k ^{\prime}|^2
+q_{TF}^2)}\langle z_{n \vec k}|
z_{n^{\prime}\vec k ^{\prime}}\rangle\nonumber.
\label{mnelement_anti}
\end{eqnarray}
Here $\epsilon_{host}$ is the host semiconductor dielectric constant,
$|z_{n \vec k}\rangle$ is the six-component envelope-function eigenspinor 
of the Hamiltonian $\hat{H}^h$,
and the Thomas-Fermi screening wavevector  
$q_{TF}=\sqrt{g(E_F)e^2/(2 \epsilon_{host}\epsilon_0)}$, 
where $g(E_F)$ is the density of states at the Fermi energy,
$E_F$.
The  inter-band scattering broadening $\Gamma_{n,n'}$
in Eq. \ref{ac_sig_AHE_dis} is then calculated by averaging 
$\Gamma_{n,n'}(\vec{k})\equiv(\Gamma_{n}(\vec{k})+\Gamma_{n'}(\vec{k}))/2$
over the allowed transitions between bands $n$ and $n'$
as in Ref. \onlinecite{sinova_prb02}.

\section{4-band spherical model}
\label{4band}

In  this  section we  briefly  summarize  an  analytic calculation  of
$\sigma_{xy}(\omega)$ for a disorder  free 4-band model with isotropic
bands,  the so-called  spherical  model.  This  model  is realized  by
taking    the   spin-orbit   coupling    to   infinity    and   taking
$\gamma_2=\gamma_3$  (equal to  2.5 for  GaAs) in  the Kohn-Luttinger
6-band model of Eq. \ref{hl}. This yields
\begin{equation}
\hat{H}^{L-4b}=\frac{\hbar^2}{2m_0}\left[
(\gamma_1+\frac{5}{2}\gamma_2)k^2-2\gamma_2(\vec            k\cdot{\vec
j})^2\right],
\label{H_4b_sph}
\end{equation}
with the anti-ferromagnetic coupling between the localized moments and
the         holes          given         as         before         by,
$h\hat{s}_z=(h/3)\hat{j}_z$.     From    the
Hamiltonian in Eq.  \ref{H_4b_sph} one can immediately see  one of the
consequences  of  a strong  spin-orbit  coupling:  for  a Bloch  state
labeled by  wavevector $\vec{k}$, the spin quantization  axis at $h=0$
is  parallel to $\vec{k}$.   It is  possible to  evaluate $\sigma^{\rm
4b}_{xy}(\omega)$  from Eq.  \ref{ac_sig_AH}  in this  model to  first
order in $h$ by completing  a straightforward but lengthy exercise in
degenerate  perturbation  theory.  This  calculation  is described  in
greater detail  in appendix A.  \cite{note3} Here we simply  state the
final result:
\begin{equation}
\sigma^{\rm 4b}_{xy}(\omega)=\int_{-\infty}^\infty 
d\omega'\frac{A_{xy}(\omega')}{\omega-\omega'},
\label{sig_4b}
\end{equation}
where the spectral function $A_{xy}(\omega)$ is 
given by different expressions in three different energy intervals. 
 For $m_{lh}E_F/\mu-(h/6)m_{lh}/m_{hh}-h/2$$<$$\omega$
$<$$m_{lh}E_F/\mu+(h/6)m_{lh}/m_{hh}+h/2$, 
\begin{eqnarray}
&A_{xy}(\omega)
= -\frac{e^2\sqrt{2\mu\omega/\hbar}}{(2\pi)^2\hbar}\left[
\left.\left(\frac{3}{8}u^2 +\frac{h }{4\hbar \omega} 
[ \frac{7}{6}u^3-2u ]\right)\right|_{1-\Delta_+}^1
\right.\nonumber\\&\left.
+\left.\left(\frac{u}{8}\left(\sqrt{1-\frac{3}{4}u^2}+
\frac{3}{2}u\right)-\frac{\sqrt{3}}{12}
\arcsin(\sqrt{3}u/2)\right)\right|_{1-\Delta_-}^{1-\Delta_+}
\right.  \nonumber\\
&\left.+\left.\frac{h }{4\hbar \omega}
\left(-u+\frac{7u^3}{12}+\sqrt{1-\frac{3}{4}u^2}\left(\frac{7}{27}
+\frac{13u^2}{18}\right)
\right)\right|_{1-\Delta_-}^{1-\Delta_+} \right]
\label{Aomlow}
\end{eqnarray}
with
\begin{eqnarray}
\Delta_\pm=\frac{\frac{h}{2}(1+\frac{\xi^2}{3})
+\hbar\tilde\omega\mp\frac{\xi}{3}\sqrt{
h^2(1+\frac{\xi^2}{3})-3\hbar^2\tilde\omega^2}}
{\frac{h}{2}(1+\frac{\xi^2}{3})},
\end{eqnarray}
$m_{hh}\equiv m_0/(\gamma_1-2\gamma_2)$, 
$m_{lh}\equiv m_0/(\gamma_1+2\gamma_2)$, 
$\xi\equiv m_{lh}/m_{lh}$,
$\mu\equiv m_{hh}m_{lh}/(m_{hh}+m_{lh})$, 
and $\tilde{\omega}=\omega-\hbar k_{lh}^2/2\mu$,
where $k_{lh}$ is the light-hole band Fermi wave-vector 
in zero exchange field.
For $m_{lh}E_F/\mu+(h/6)m_{lh}/m_{hh}+h/2$
$<$$\omega$$<$$m_{hh}E_F/\mu-h/6-(h/2)m_{hh}/m_{lh}$
\begin{eqnarray}
A_{xy}(\omega)=
\frac{e^2}{(2\pi\hbar)}\frac{5}{24\pi}
\sqrt{\frac{2\mu\hbar\omega}{\hbar^2}}
\frac{h}{\hbar\omega}.
\label{Aommed}
\end{eqnarray}
For $m_{hh}E_F/\mu-h/6-(h/2)m_{hh}/m_{lh}$$<$$\omega$$<$
$m_{hh}E_F/\mu+h/6+(h/2)m_{hh}/m_{lh}$,
\begin{eqnarray}
& A_{xy}(\omega)=-\frac{e^2\sqrt{2\mu\omega/\hbar}}
{(2\pi)^2\hbar}\left[
\left.\left(\frac{3}{8}u^2 +\frac{h }{4\hbar \omega} 
[ \frac{7}{6}u^3-2u ]\right)\right|^{1-\tilde\Delta_+}_{-1}
\right.\nonumber\\
& \left.
+\left.\left(\frac{u}{8}\left(\sqrt{1-\frac{3}{4}u^2}+
\frac{3}{2}u\right)-\frac{\sqrt{3}}{12}
\arcsin(\sqrt{3}u/2)\right)\right|^{1-\tilde\Delta_-}_{1-
\tilde\Delta_+}\right.  \nonumber\\
&\left.+\left.\frac{h }{4\hbar \omega}
\left(\frac{7u^3-12 u}{12}+\sqrt{\frac{4-4u^2}{4}}
\left(\frac{126+351 u^2}{486}\right)
\right)\right|^{1-\tilde\Delta_-}_{1-\tilde\Delta_+} \right]
\label{Aomhigh}
\end{eqnarray}
with
\begin{eqnarray}
\tilde\Delta_{\pm}=\frac{\frac{h}{2}(1+\frac{\xi^2}{3})+\xi\hbar\tilde
\omega\mp\frac{\xi}{3}\sqrt{
h^2(1+\frac{\xi^2}{3})-3\xi^2\hbar^2\tilde\omega^2}}
{\frac{h}{2}(1+\frac{\xi^2}{3})},
\label{tildeDelta}
\end{eqnarray} 
and $\tilde{\omega}=\omega-\hbar k_{hh}^2/2\mu$,
where $k_{hh}$ is the heavy-hole band Fermi 
wave-vector in zero exchange field;
and $A_{xy}(\omega)=0$ otherwise.
\begin{figure}
\includegraphics[width=3.4in]{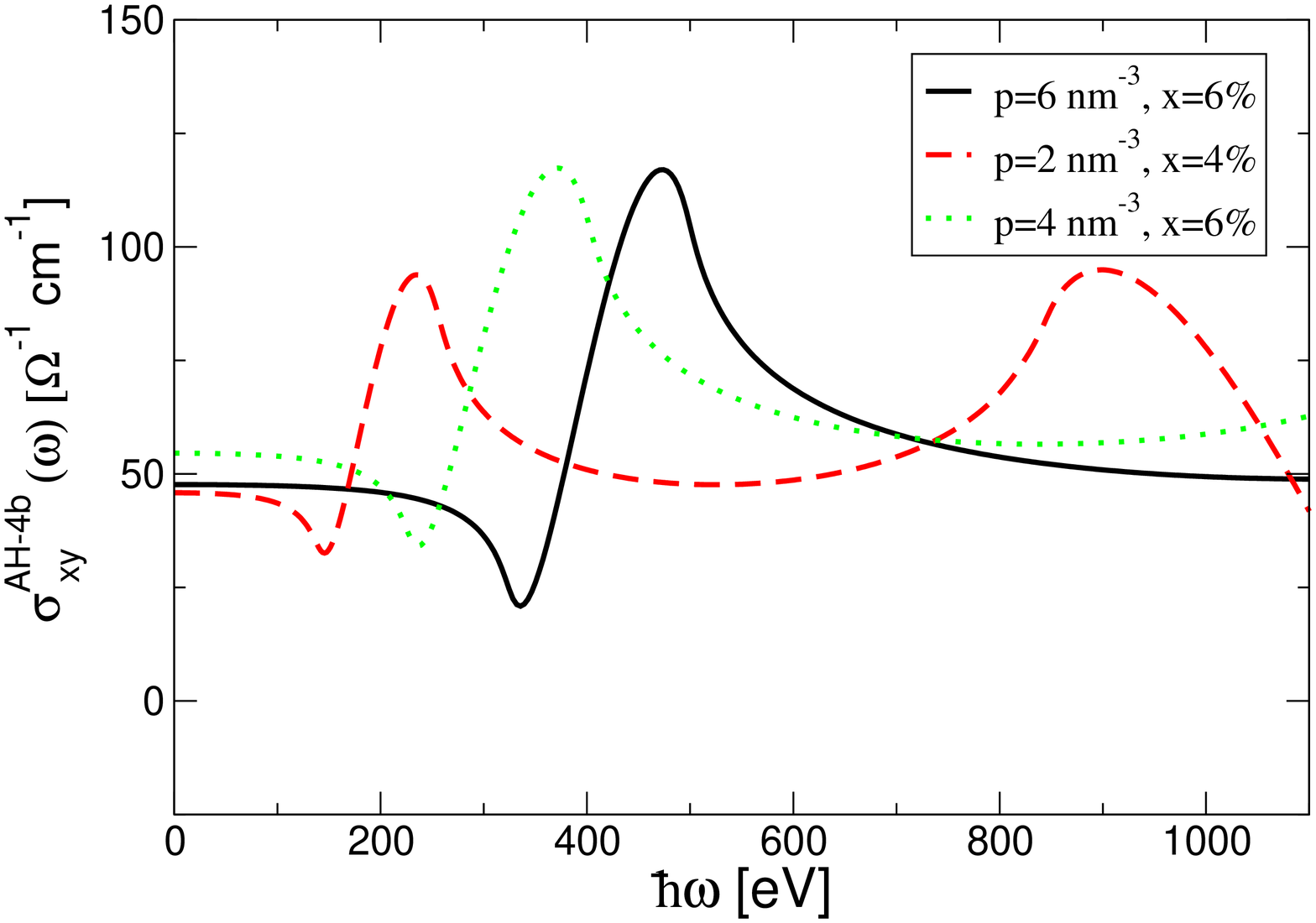}
\caption{Anomalous ac-Hall conductivity calculated within
the 4-band spherical model without disorder life-time broadening
for several itinerant hole and Mn concentrations.}
\label{fig4b}
\end{figure}
% Comment: Add a schematic figure here and tie in with these expressions.
% Most importantly, identify which types of interband transitions are contributing in the 
% three different energy intervals above.  Add a paragraph of dicussion of the 
% figure with the aim of really generating some intuition about why the 
% expressions look the way they do.

\begin{figure}
\includegraphics[width=3.4in]{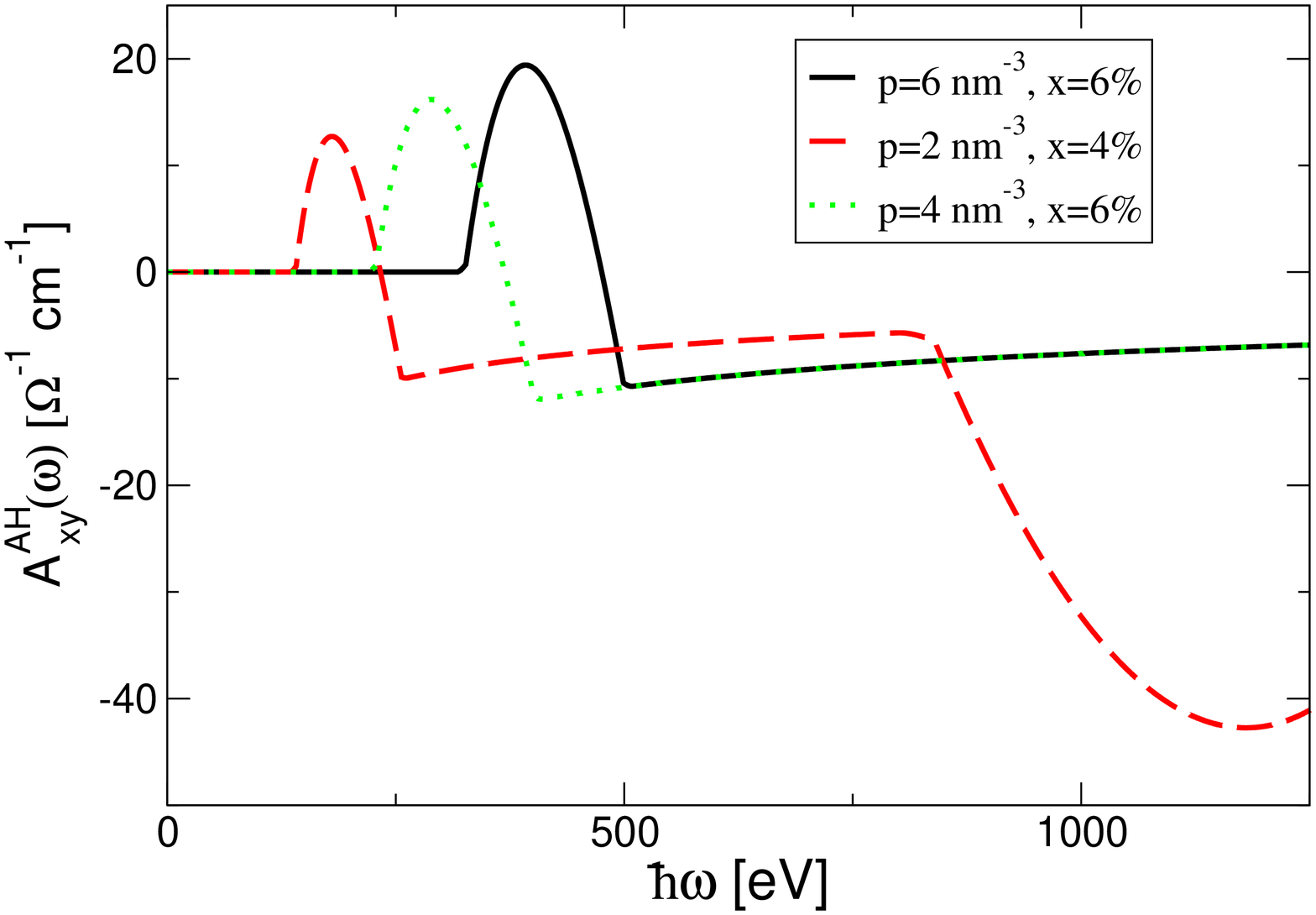}
\caption{Spectral  function  $A_{xy}(\omega)$  calculated  within  the
4-band   model   the  itinerant   hole   and   Mn  concentrations   of
Fig. \ref{fig4b}.}
\label{Aomega}
\end{figure}

We show  $\sigma^{4b}_{xy}(\omega)$ for several itinerant  hole and Mn
concentrations in  Fig. \ref{fig4b}.  From the above  result (and from
the details  presented in Appendix A)  it is relatively  simple to see
the source  of the  feature observed in  the mid-infrared  regime. The
spectral function $A_{xy}(\omega)$, shown in Fig. \ref{Aomega} for the
% Comment: Tie this discussion to the new figure? 
parameters used  in Fig. \ref{fig4b}, has its  major contribution from
transitions near  the light-holes bands Fermi  wave-vectors (the lower
frequency  peak in  $A_{xy}(\omega)$) and  near the  heavy-holes Fermi
wave-vectors (the higher  frequency peak in $A_{xy}(\omega)$), visible
for $x=4\%$ and $p=0.2 {\rm nm}^{-3}$. The transitions that contribute
to first order in $h$ are  between heavy and light holes with opposite
polarization as shown in Appendix \ref{4bdetails}.
We also note that there is a considerable contribution to $\sigma^{\rm
4b}_{xy}(\omega)$  from  the  high  frequency  part  of  the  spectral
function (accounts for rigid shifts  in the low frequency range) which
indicate the  possible need to consider higher  bands, maybe including
the conduction bands, for more realistic calculations.

\section{Numerical Results}
\label{6band}
The qualitative physics behind the the 4-band model calculation
results still applies to the full model numerical calculations. However, the
effects on $\sigma_{xy}(\omega)$ due to the 
lifetime broadening of the quasiparticles, finite spin-orbit coupling,
and the warping of the bands ($\gamma_3 \ne\gamma_2$) 
at higher concentrations are an important part of the 
quantitative numerical result.
\begin{figure}
\includegraphics[width=3.4in]{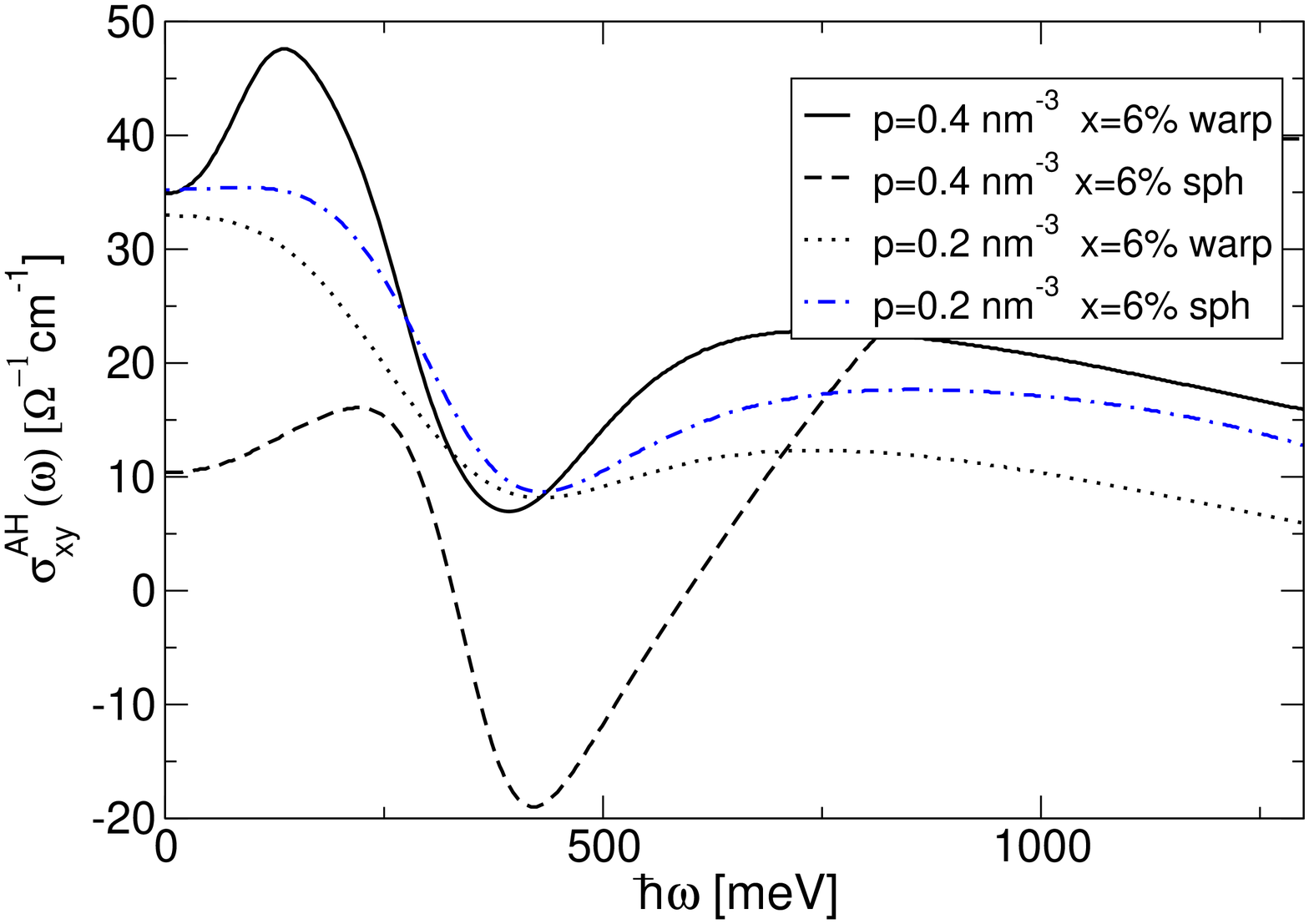}
\caption{Anomalous ac-Hall conductivity $\sigma_{xy}(\omega)$
for $x=6\%$ Mn concentration and $p=0.4$ and $0.2 {\rm nm}^{-3}$,
for spherical and non-spherical (band-warping) models.}
\label{sigAH_x6_p4}
\end{figure}
As an example, Fig. \ref{sigAH_x6_p4} shows the anomalous ac-Hall conductivity
of disordered system 
for $x=6$ \% and $p=0.2$ and $0.4$ nm$^{-3}$ calculated
using the 6-band model with warping ($\gamma_3 \ne\gamma_2$)
and without 
warping ($\gamma_3 =\gamma_2$), 
which emphasizes the importance of including the warping
of the bands in obtaining reliable results which can be compared
directly with experiment. 

The Hall conductivity 
must be non-zero in order to have non-zero magneto-optical effects, but most
measurable quantities are also influenced by other elements of the 
conductivity tensor.
The most widely studied magneto-optical effects
are the Faraday and Kerr effects.  The Faraday effect reflects 
the relative difference between the optical absorption of
right and left circularly polarized light, 
referred to as magnetic circular dichroism (MCD).
In the Voigt geometry (magnetization aligned with
axis of light propagation) and assuming a thin film geometry (applicable
for all (III,Mn)V epilayers now available in the infrared regime 
considered here)
\begin{equation}
MCD=\frac{\alpha^+-\alpha^-}{\alpha^++\alpha^-}=\frac{{\rm Im}[\sigma_{xy}(\omega)]}
{{\rm Re}[\sigma_{xx}(\omega)]}.
\end{equation}
Linearly polarized light propagating through a magnetic medium will
experience the Faraday rotation of its polarization angle and a transformation 
from linear to elliptically polarized light due to MCD. The angle
of rotation per unit length traversed, again in the thin film geometry, is
(in cgs units)
\begin{equation}
\theta_F(\omega)=\frac{4\pi}{(1+n)c}{\rm Re}[\sigma_{xy}],
\end{equation}
where $c$ is the speed of light and $n$ is the index of refraction of the
substrate, in this case GaAs with $n=\sqrt{10.9}$.
Perhaps the more technologically relevant magneto-optic phenomena is the Kerr
effect, which appears in reflection from a magnetic medium. In this
case, also within the Voigt geometry, the Kerr angle and ellipticity are defined
as
\begin{equation}
\theta_K+i\eta_K\equiv\frac{r_+-r_-}{r_++r_-},
\end{equation}
where $r_\pm$ are the total complex reflection amplitudes (with multiple scattering 
taken into account) for right and left circular polarized light. 
Note that the simple relations,
$\theta_K\propto {\rm Im}[\sigma_{xy}(\omega)]$
and $\eta_K\propto {\rm Re}[\sigma_{xy}(\omega)]$, \cite{ando_book}
obtained in the thick-layer limit do not apply for the typical thin (III,Mn)V
epilayers.
\begin{figure}
\includegraphics[width=3.2in]{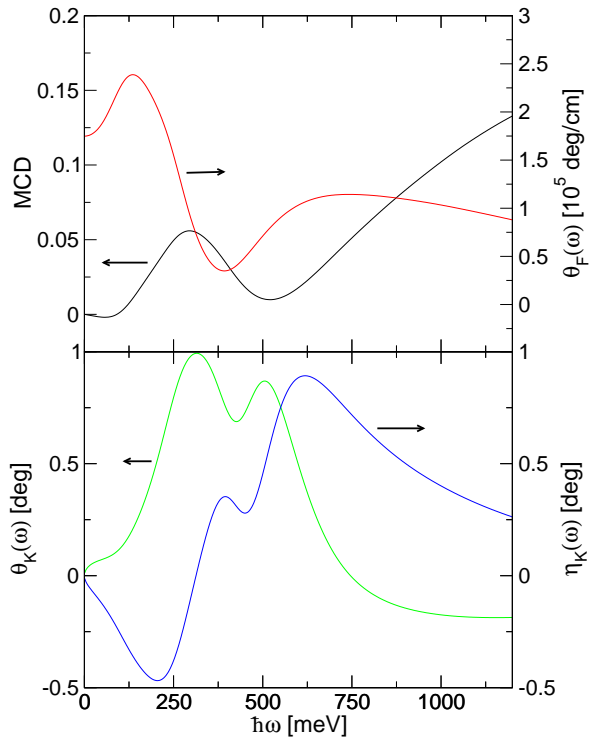}
\caption{Faraday and Kerr effects 
for $x=6\%$ Mn concentration and $p=0.4$ ${\rm nm}^{-3}$.}
\label{mo_eff_x6_p4}
\end{figure}
In Fig. \ref{mo_eff_x6_p4} we show the different magneto-optic effects for a 
concentration of $x=6\%$ and $p=0.4$ ${\rm nm}^{-3}$. The Faraday rotation
in this case is larger than the giant Faraday rotation observed in the
paramagnetic (II,Mn)VI's at optical frequencies \cite{ando_book,koyanagi_jap87} and
should be readily observable in the current highly metallic samples.
The Kerr angle and ellipticity we obtain for (Ga,Mn)As are 
comparable to the Kerr effects observed
in the optical regime in materials used for magneto-recording devices. 
\cite{kaneko_book}
The behavior as a function of free carrier hole concentration can be seen
in Fig. \ref{faraday_x6} where the Faraday rotation angle is shown for several 
carrier concentrations. 
The peaks and valleys in the different quantities are present 
in all the concentrations, however the magnitude
varies, even changing sign at several concentrations and frequencies.
\begin{figure}
\includegraphics[width=3.4in]{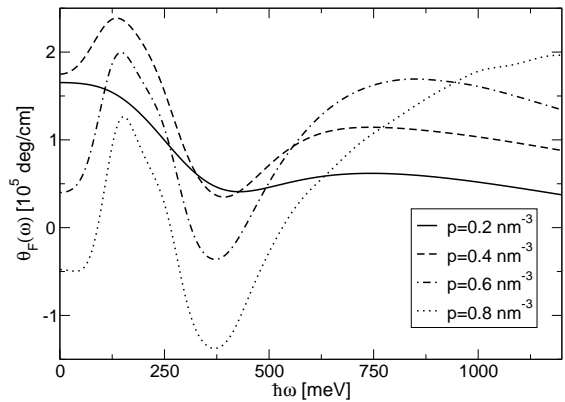}
\caption{Faraday rotation angle 
for $x=6\%$ Mn concentration with $p=0.2, 0.4, 0.6,$ 
and $0.8$  ${\rm nm}^{-3}$.}
\label{faraday_x6}
\end{figure}

Rather  than presenting  many different  graphs for  all  the possible
parameters  ($p$,$x$,  etc.), we  direct  the  reader  to a  data-base
located   at   http://unix12.fzu.cz/ms,   where  results   for   these
quantities, together  with other physical quantities,  can be obtained
and plotted vs. different nominal parameters.\cite{note2}

\section{Conclusions}
\label{conclusion}

We  have presented  a theory  of the  ac Hall  effect in  the infrared
regime by extending the Berry's  phase theory of the dc-anomalous Hall
effect  to finite frequencies  and treating the
effects  of disorder  through a  finite lifetime  of  the valence-band
quasiparticles.   We  observe 
features  (peaks and valleys)  in the  transverse conductivity  in the
range between 200 and 400 meV at which the conductivity changes by 
more than  100\%.   We have studied how these features appear in 
different magneto-optical effects (MCD, Faraday rotation and Kerr effect)
which are relatively easily measured, finding strong  signals.
The magnitude  of the Faraday rotation is very
large  (one order of  magnitude larger than that observed  in 
paramagnetic (II,Mn)VI's for example) and has a nontrivial dependence on the free
carrier concentration.  The Kerr effect is also strong
when compared to materials used in magneto-optic recording.   
The origin of the peaks is most easily 
understood within  a simple 4-band spherical model in which transitions
between heavy and light holes states with opposite spin-polarization 
give the strongest contribution to the
anomalous transverse optical conductivity.  The four band model represents
the infinite spin-orbit coupling strength limit of the six-band model we use for 
numerical calculations.  Our use of a six-band model can account only
for transitions within the valence band and not for transitions between 
conduction and valence bands.  Because of this limitation, we cannot address
the crossover between intra-band and interband contributions which are not
completely separated in these extremely heavily-doped semiconductors, something
that is clearly desirable and should be addressed in subsequent theoretical work.

Our predictions depend in intricate detail on the model that we have used
to describe the ferromagnetism of these materials.  The model depends 
most essentially on the assumption that the Mn impurities act as reasonably 
shallow acceptors and introduce $S=5/2$ local moment degrees of freedom to 
the system.  The specific calculations presented here assume that 
Mn impurities and other scatterers in the system can be treated perturbatively.  
This assumption enables quasiparticle scattering rates to be estimated in a 
simple way, but is a less essential part of the model.  The 
magneto-optical properties studied here are directly dependent on 
valence-band spin-orbit coupling, which we have argued elsewhere \cite{bookchapter,
koenig_prb2001} plays an essential role in understanding ferromagnetism in
these materials.  Confirmation by future experiment of the 
detailed predictions made here for the magneto-optical properties
of these materials would further validate the approach we have taken
to modeling these interesting new ferromagnets. 
We expect that the weak-quasiparticle-scattering approximations made 
here will be more reliable in more metallic samples, since  
the scattering rates are then smaller compared to other relevant energy scales,
particularly the Fermi  energy.  We hope that these calculations will help 
motivate magneto-optic experiments  in the  infrared  regime for 
(Ga,Mn)As and other (III,Mn)V ferromagnets. 

\begin{acknowledgments}

The authors gratefully acknowledge stimulating conversations
with D. Basov, B. Gallagher, T. Dietl, and Q. Niu.
This work was supported by the Welch Foundation, 
by the Office of Naval Research under grant N000140010951, 
and by the Grant Agency of the Czech Republic under grant 202/02/0912.

\end{acknowledgments}

\appendix

\section{Derivation of $\sigma_{xy}(\omega)$ in the 4-band spherical model}
\label{4bdetails}
We present in this appendix the details involved in deriving the results
shown in Eqs. \ref{sig_4b}-\ref{tildeDelta} for
the anomalous contribution to the ac Hall conductivity 
calculated first in the exchange field within 
the 4-band spherical model.  The host valence band Hamiltonian in this case,
as shown in Sec. \ref{4band}, is given by
\begin{equation}
\hat{H}^{L-4b}=\frac{\hbar^2}{2m_0}\left[
(\gamma_1+\frac{5}{2}\gamma_2)k^2-2\gamma_2(\kk\cdot{\bf j})^2\right],
\end{equation}
The  eigenspinors of $\hat{H}^{L-4b}$ are given by
\begin{equation}
|z_{nk}^{(0)}\rangle=e^{-i{\hat{j}_z\phi}/{\hbar}}
e^{-i{\hat{j}_y\theta}/{\hbar}}|n\rangle,
\label{z_0}
\end{equation}
where $|n\rangle$ are the spinors with the axis of quantization along the
z-direction and total angular momentum $3/2\hbar$. 
The perturbation due to the antiferromagnetic
coupling to the localized moments is $\hat{H}'=h\hat{s}_z=(h/3)\hat{j}_z$. 
The eigenvalues to first order in $h$ are then given by
\begin{equation}
E_{hh}^{\pm}=\frac{\hbar^2k^2}{2m_{hh}}\pm\frac{h}{2}\cos\theta
\end{equation}
and
\begin{equation}
E_{lh}^{\pm}=\frac{\hbar^2k^2}{2m_{lh}}\pm\frac{h}{3}\sqrt{1-\frac{3}{4}\cos^2\theta}
=\frac{\hbar^2k^2}{2m_{lh}}\pm\frac{h}{6}\frac{\cos\theta}{\cos 2\theta'},
\end{equation}
where $\tan 2\theta'=2\tan\theta$, $hh$ labels heavy-holes and $lh$ 
labels light-holes.
\begin{widetext}
The dipole matrix elements in Eq. \ref{ac_sig_AH} are given by:
\begin{eqnarray}
\langle n' k|\hat{p}_\alpha|nk\rangle=\frac{m}{\hbar}
\langle \znpk|\frac{\partial H}{\partial k_\alpha}|\znk\rangle=
\frac{m(E_{nk}-E_{n'k})}{\hbar}\langle\frac{\partial}{\partial k_\alpha}
n' k|nk\rangle,
\label{dipole1}
\end{eqnarray}
so we can write
\begin{eqnarray}
{\rm Im}[\langle n' k|\hat{p}_x|nk\rangle\langle nk|\hat{p}_y|n' k\rangle]=
\frac{m^2}{\hbar^2}(E_{nk}-E_{n'k})^2
{\rm Im}\left[\langle z_{n'k}|\frac{\partial}{\partial k_x}\znk\rangle
\overline{\langle z_{n'k}|\frac{\partial}{\partial k_y}\znk\rangle}
\right],
\label{dipole2}
\end{eqnarray}
where
\begin{eqnarray}
\left|\frac{\partial}{\partial k_x}\znk\rangle\right.
&=&\frac{\cos\phi\cos\theta}{k}\frac{\partial}{\partial\theta}|\znk\rangle
-\frac{\sin\phi}{k\sin\theta}\frac{\partial}{\partial\phi}|\znk\rangle
+\cos\phi\sin\theta\frac{\partial}{\partial k}|\znk\rangle,
\label{der_zx}
\end{eqnarray}
and similarly
\begin{eqnarray}
\left|\frac{\partial}{\partial k_y}\znk\rangle\right.=
\frac{\sin\phi\cos\theta}{k}\frac{\partial}{\partial\theta}|\znk\rangle
-\frac{\cos\phi}{k\sin\theta}\frac{\partial}{\partial\phi}|\znk\rangle
+\sin\phi\sin\theta\frac{\partial}{\partial k}|\znk\rangle.
\label{der_zy}
\end{eqnarray}
The perturbed spinor wave function can be written as
\begin{equation}
|z_{nk}\rangle=\sum_{n'}C^n_{n'}(\theta,k)|z_{n'k}^{(0)}\rangle=
\sum_{n'}C^n_{n'}(\theta,k)e^{-i{(\hat{j}_z-j^n(0))\phi}/{\hbar}}
e^{-i{\hat{j}_y\theta}/{\hbar}}|n'\rangle
\equiv |\tilde{z}_{nk}\rangle-\frac{i}{\hbar}(\cos\phi \hat{j}_y
-\sin\phi \hat{j}_x)|n'\rangle,
\label{zz}
\end{equation}
where $j^n(0)\equiv \langle z_{n\kk=k\hat{z}}|\hat{j}_z|z_{n\kk=k\hat{z}}\rangle$. 
Inserting Eq. \ref{zz} into Eq. \ref{der_zx} and \ref{der_zy} gives
\begin{eqnarray}
|\frac{\partial}{\partial k_x}\znk\rangle&=&
\frac{i\sin\phi}{\hbar k\sin\theta}(\hat{j}_z-j^n(0))|z_{nk}\rangle
-i\frac{\cos\phi\cos\theta}{\hbar k}\left(
[\cos\phi \hat{j}_y-\sin\phi\hat{j}_x]|z_{nk}\rangle+i |\tilde{z_{nk}}\rangle\right)
+\cos\phi\sin\theta\frac{\partial}{\partial k}|\znk\rangle\label{pzkx},\nonumber\\
|\frac{\partial}{\partial k_y}\znk\rangle&=&
\frac{-i\cos\phi}{\hbar k\sin\theta}(\hat{j}_z-j^n(0))|z_{nk}\rangle
-i\frac{\sin\phi\cos\theta}{\hbar k}\left(
[\cos\phi \hat{j}_y-\sin\phi\hat{j}_x]|z_{nk}\rangle+i |\tilde{z_{nk}}\rangle\right)
+\sin\phi\sin\theta\frac{\partial}{\partial k}|\znk\rangle\label{pzky}\nonumber,
\end{eqnarray}
which can be inserted in Eq. \ref{dipole2} to yield
\begin{eqnarray}
{\rm Im}[\langle n' k|\hat{p}_x|nk\rangle\langle nk| \hat{p}_y|n' k\rangle]&=&
\frac{m^2}{\hbar^2}{(E_{nk}-E_{n'k})^2}
\langle z_{n'k}|(\hat{j}_z-j^n(0))|z_{nk}\rangle
{\rm Im}\left[
\frac{\cos\theta}{(\hbar k)^2\sin\theta}
(\langle z_{n'k}|(\cos\phi \hat{j}_y-\sin\phi\hat{j}_x)|z_{nk}\rangle\right.\nonumber\\
&&\left. +i \hbar \langle z_{n'k}|\tilde{z_{nk}}\rangle
+\frac{i}{\hbar k}\langle z_{n'k}|\frac{\partial z_{nk}}{\partial k}\rangle\right],
\end{eqnarray}
where
\begin{eqnarray}
\langle z_{n'k}|(\hat{j}_z-j^n(0))|z_{nk}\rangle&=&
\sum_{n_1 n_2} C^{n'}_{n_1}(\theta,k)C^{n}_{n_2}(\theta,k)\langle n_1|(\hat{j}_z-j^n(0))
|n_2\rangle,\nonumber\\
{\rm Im}\left[\langle z_{n'k}|(\cos\phi \hat{j}_y-\sin\phi\hat{j}_x)|z_{nk} \rangle\right]&=&
\sum_{n_1 n_2} C^{n'}_{n_1}(\theta,k)C^{n}_{n_2}(\theta)\langle n_1|\hat{j}_y|n_2\rangle,\nonumber\\
\langle z_{n'k}|\tilde{z_{nk}}\rangle=\sum_{n_1}C^{n'}_{n_1}(\theta,k)\frac{\partial
C^{n}_{n_1}(\theta,k)}{\partial\theta}\,\,\,&{\rm and}&\,\,\,
\langle z_{n'k}|\frac{\partial z_{nk}}{\partial k}\rangle=
\sum_{n_1}C^{n'}_{n_1}(\theta,k)\frac{\partial C^{n}_{n_1}(\theta,k)}
{\partial k}\nonumber.
\end{eqnarray}

Here we only need to consider six transitions since we only need the 
$n\ne n'$ terms and we will ignore transitions between bands with equal 
effective masses which can be shown to contribute to higher 
order in $h$.  From degenerate perturbation theory we obtain the four 
eigenvectors to linear order in $h$:
\begin{eqnarray}
|\kk,hh\pm\rangle&=&|\kk,\pm3/2\rangle+\frac{h \mu \sin\theta}
{\sqrt{3}(\hbar k)^2}|\kk,\pm 1/2\rangle,\\\nonumber\\
|\kk,lh+\rangle&=&\cos{\theta'}|\kk, +1/2\rangle-\sin{\theta'}|\kk, -1/2\rangle
-\frac{h \mu \sin\theta}{\sqrt{3}(\hbar k)^2}
[\cos{\theta'}|\kk, +3/2\rangle-\sin{\theta'}|\kk, -3/2\rangle],\\\nonumber\\
|\kk,lh-\rangle&=&\sin{\theta'}|\kk, +1/2\rangle+\cos{\theta'}|\kk, -1/2\rangle
-\frac{h \mu \sin\theta}{\sqrt{3}(\hbar k)^2}
[\sin{\theta'}|\kk, +3/2\rangle+\cos{\theta'}|\kk, -3/2\rangle],
\end{eqnarray}
where $\mu\equiv{m_{lh}m_{hh}}/({m_{hh}-m_{lh}})$.
The Fermi wavevectors to first order in $h/E_F$  for each band are given by
\begin{eqnarray}
k_F^{hh\pm}(\theta)=k_F^{hh(0)}\left(1\pm\frac{h}{4E_F}\cos\theta\right)\,\,\,{\rm and}\,\,\,
k_F^{lh\pm}(\theta)=k_F^{lh(0)}\left(1\pm\frac{h}{6E_F}\sqrt{1-\frac{3}{4}\cos^2\theta}\right).
\end{eqnarray}

After some lengthy algebra one obtains
\begin{eqnarray}
\frac{{\rm Im}[\langle k,hh+|\hat{p}_x|k,lh+\rangle\langle k,lh+| 
\hat{p}_y|k,hh+\rangle]}
{(E_{lh}^+-E_{hh}^+)}
&=& \frac{3m^2}{8\mu}\cos\theta\cos^2\theta'+\frac{h m^2}{2(\hbar k)^2}
(-\frac{1}{4}\sin(2\theta)\sin(2\theta')
+\cos 2\theta\cos^2\theta'\nonumber\\&&
+\frac{\cos^2\theta\cos^2\theta'}{4\cos 2\theta'}
-\frac{3}{4}\cos^2\theta\cos^2\theta')\label{d_hplp},\\
\frac{{\rm Im}[\langle k,hh+|\hat{p}_x|k,lh-\rangle\langle k,lh-| 
\hat{p}_y|k,hh+\rangle]}
{(E_{hl}^--E_{hh}^+)}
&=&\frac{3m^2}{8\mu}\cos\theta\sin^2\theta'+\frac{h m^2}{2(\hbar k)^2}
(+\frac{1}{4}\sin(2\theta)\sin(2\theta')+\cos 2\theta\sin^2\theta'
\nonumber\\&&-\frac{\cos^2\theta\sin^2\theta'}{4\cos 2\theta'}
-\frac{3}{4}\cos^2\theta\sin^2\theta')\label{d_hplm},\\
\frac{{\rm Im}[\langle k,hh-|\hat{p}_x|k,lh+\rangle\langle k,lh+| 
\hat{p}_y|k,hh-\rangle]}
{(E_{hl}^+-E_{hh}^-)} &=&-\frac{3m^2}{8\mu}\cos\theta\sin^2\theta'
+\frac{h m^2}{2(\hbar k)^2}
(+\frac{1}{4}\sin(2\theta)\sin(2\theta') +\cos 2\theta\sin^2\theta'\nonumber\\
&&-\frac{\cos^2\theta\sin^2\theta'}{4\cos 2\theta'}
-\frac{3}{4}\cos^2\theta\sin^2\theta')\label{d_hmlp},\\
\frac{{\rm Im}[\langle k,hh-|\hat{p}_x|k,lh-\rangle\langle k,lh-| \hat{p}_y|k,hh-\rangle]}
{(E_{lh}^--E_{hh}^-)}
&=&-\frac{3m^2}{8\mu}\cos\theta\cos^2\theta'+\frac{h m^2}{2(\hbar k)^2}
(-\frac{1}{4}\sin(2\theta)\sin(2\theta')+\cos 2\theta\cos^2\theta')\nonumber\\
&&+\frac{\cos^2\theta\cos^2\theta'}{4\cos 2\theta'}
-\frac{3}{4}\cos^2\theta\cos^2\theta')\label{d_hmlm}.
\end{eqnarray}

Using Eqs. \ref{d_hplp}-\ref{d_hmlm} we can compute directly the 
dc conductivity (Eq. \ref{ac_sig_AH} for $\omega=0$): 
\begin{eqnarray}
\sigma_{xy}(0)&=&\frac{2e^2\hbar}{m^2 V}\sum_{\kk,n> n'}
\frac{(f_{n',k}-f_{n,k})
{\rm Im}[\langle n' k|\hat{p}_x|nk\rangle\langle nk| \hat{p}_y|n' k\rangle]}
{(E_{nk}-E_{n'k})^2}\nonumber\\
&=&-\frac{e^2}{(2\pi\hbar)}\frac{hk_F^{hh0}}{4\pi E_F}\left[1-\frac{1}{3}\sqrt{\frac{m_{lh}}{m_{hh}}}
+\frac{8}{3}\frac{m_{lh}}{m_{lh}+\sqrt{m_{lh}m_{hh}}}\right],
\end{eqnarray}
in agreement with the previously derived dc-anomalous Hall conductivity 
\cite{jungwirth_prl02} using the Berry's phase contribution to the
Bloch group velocity in the semi-classical equations of motion approach.

To compute the ac-anomalous Hall conductivity given by
Eq. \ref{ac_sig_AH} we rewrite it in terms of the spectral function 
$A_{xy}(\omega)$
\begin{equation}
\sigma_{xy}(\omega)=\int_{-\infty}^\infty 
d\omega'\frac{A_{xy}(\omega')}{\omega-\omega'},
\end{equation}
with
\begin{equation}
A_{xy}(\omega)\equiv-\frac{e^2\hbar}{m^2 V}\sum_{\kk,n\ne n'}
\frac{(f_{n',k}-f_{n,k})
{\rm Im}[\langle n' k|\hat{p}_x|nk\rangle\langle nk| \hat{p}_y|n' k\rangle]}
{(E_{nk}-E_{n'k})}\delta(\hbar\omega-(E_{nk}-E_{n'k})).
\end{equation}
$A_{xy}(\omega)$ is an odd function of $\omega$  and we need only to consider
$\omega>0$. We need to consider three separate frequency ranges in what follows.
First we look at the range
\begin{equation}
\frac{m_{lh} E_F}{\mu}+ \frac{m_{lh}}{m_{hh}}\frac{h}{6}+\frac{h}{2}<\omega
<\frac{m_{hh} E_F}{\mu}- \frac{h}{6}-\frac{m_{hh}}{m_{lh}}\frac{h}{2},
\end{equation}
and consider the different contributions to $A_{xy}(\omega)$ from the
four types of transitions, $hh^\pm\rightarrow lh^\pm$, separately.
For $hh+$ to $lh\pm$ transitions we have
\begin{eqnarray*}
&&A_{xy}(\omega;hh+\rightarrow lh\pm)\nonumber\\&=&\left.
-\frac{e^2\hbar }{m^2 (2\pi)^2}\int_{-1}^1d(\cos\theta)\frac{\mu k}{\hbar^2}
\frac{{\rm Im}[\langle k,hh+|\hat{p}_x|k,lh\pm\rangle\langle k,lh\pm| \hat{p}_y|k,hh+\rangle]}
{(E_{hl}^\pm-E_{hh}^+)}
\right|_{k=\sqrt{\frac{2\mu\omega}{\hbar}}\left(1\mp\frac{h\cos\theta}{12\hbar\omega\cos2\theta'}
+\frac{h}{4\hbar\omega}\cos\theta\right)}\\
&=& -\frac{e^2\sqrt{2\mu\omega/\hbar}}{(2\pi)^2\hbar}\int_{-1}^1d(\cos\theta)
\left(\frac{3}{8}\cos\theta
\left\{\begin{array}{c} \cos^2\theta'\\\sin^2\theta' \end{array}\right\}
+\frac{h }{4\hbar \omega} [\mp\frac{1}{4}\sin(2\theta)\sin(2\theta')
+\cos 2\theta\left\{\begin{array}{c} \cos^2\theta'\\\sin^2\theta' \end{array}\right\}\right.\\&&\left.
\pm\frac{1}{8}\frac{\cos^2\theta}{\cos 2\theta'}
\left\{\begin{array}{c} \cos^2\theta'\\\sin^2\theta' \end{array}\right\}
-\frac{3}{8}\cos^2\theta\left\{\begin{array}{c} \cos^2\theta'\\\sin^2\theta' \end{array} \right\}
]\right).\\
\end{eqnarray*}
We can sum the two and obtain
\begin{eqnarray}
A_{xy}(\omega;hh+\rightarrow lh+)+
A_{xy}(\omega;hh+\rightarrow lh-)&=& 
-\frac{e^2\sqrt{2\mu\omega/\hbar}}{(2\pi)^2\hbar}\int_{-1}^1d(\cos\theta)
\left(\frac{3}{8}\cos\theta +\frac{h }{4\hbar \omega} [ +\cos 2\theta
+\frac{1}{8}{\cos^2\theta} -\frac{3}{8}\cos^2\theta ]\right)\nonumber\\
&=&\frac{e^2}{(2\pi\hbar)}\frac{5}{48\pi}\sqrt{\frac{2\mu\hbar\omega}{\hbar^2}}
\frac{h}{\hbar\omega}
\end{eqnarray}
For the $hh-$ to $lh\pm$ transition we obtain the same result, 
therefore within this range
%$\frac{m_{lh}}{\mu}E_F+\frac{m_{lh}}{m_{hh}}\frac{h}{6}+\frac{h}{2}
%< \omega<\frac{m_{mh}}{\mu}E_F-\frac{h}{6}-\frac{m_{hh}}{m_{hl}}\frac{h}{2}$ 
we have
\begin{eqnarray*}
A_{xy}(\omega)=
\frac{e^2}{(2\pi\hbar)}\frac{5}{24\pi}\sqrt{\frac{2\mu\hbar\omega}{\hbar^2}}
\frac{h}{\hbar\omega}.
\end{eqnarray*}
As one can see from its definition  $A_{xy}(\omega)$ 
changes most rapidly in the region where 
transitions near the Fermi surface are allowed. 
Let's next consider transitions from $hh+$ to $lh\pm$ first
in the lower range 
$\frac{m_{lh}}{\mu}E_F-\frac{m_{lh}}{m_{hh}}\frac{h}{6}-\frac{h}{2}<\omega
\frac{m_{lh}}{\mu}E_F+\frac{m_{lh}}{m_{hh}}\frac{h}{6}+\frac{h}{2}$:
\begin{eqnarray}
A_{xy}(\omega;hh+\rightarrow lh\pm)=
\int_{-1}^{1}d(\cos\theta)\int_{k_F^{lh\pm}(\theta)}^\infty 
dk f(\theta,k) \delta(\hbar\omega-\Delta E^\pm),
\end{eqnarray}
with
\[\Delta E^\pm_+=
\frac{(\hbar k)^2}{2\mu}\pm\frac{h\cos\theta}
{6\cos 2\theta'}-\frac{h}{2}\cos\theta\,\,\,\,
{\rm and}\,\,\,\,\, 
k_F^{lh\pm}(\theta)=k_F^{lh(0)}\left(1\mp\frac{h}
{6E_F}\sqrt{1-\frac{3}{4}\cos^2\theta}\right).
\]
\end{widetext}
The minimum of $\Delta E^\pm(\theta)_+$  at a fixed $\theta$ is then
\begin{eqnarray}
\Delta E^\pm_+(\theta)&=&
\frac{\hbar^2}{2\mu}{k_F^{lh(0)}}^2\pm\xi
\frac{h\cos\theta}{6\cos 2\theta'}-\frac{h}{2}\cos\theta,
\end{eqnarray}
where we have defined $\xi=-m_{lh}/\mu+1=m_{lh}/m_{hh}$, 
and the absolute minimum is given by
\[
\Delta E^\pm_+(\theta_{min}=0)=\frac{\hbar^2}
{2\mu}{k_F^{lh(0)}}^2\pm\frac{m_{lh}}{m_{hh}}
\frac{h}{6}-\frac{h}{2}.
\]
For $hh-$ to $lh\pm$ we have instead
\begin{eqnarray}
\Delta E^\pm_-(\theta)&=&\frac{\hbar^2}{2\mu}
{k_F^{lh(0)}}^2\pm\frac{m_{lh}}{m_{hh}}
\frac{h\cos\theta}{6\cos 2\theta'}+\frac{h}{2}\cos\theta\nonumber,\\
\Delta E^\pm_-(\theta_{min}=\pi)&=&\frac{\hbar^2}
{2\mu}{k_F^{lh(0)}}^2\pm\frac{m_{lh}}{m_{hh}}
\frac{h}{6}-\frac{h}{2}\nonumber.
\end{eqnarray}
Let $\hbar\omega=\frac{\hbar^2}{2\mu}{k_F^{lh(0)}}^2+\tilde\omega$
where $\hbar\tilde\omega$ will  be of the order of $h$. 
For an $\tilde\omega$ that is too small there will be a limit on the
angular integration, $\tilde\theta$, obtained by setting $k=k_F^{lh\pm}$ so
for $hh+$ to $lh\pm$
\begin{eqnarray}
\hbar\tilde\omega&&\mp\frac{m_{lh}}{m_{hh}}\frac{h}{3 }
\sqrt{1-\frac{3}{4}\cos^2\tilde\theta}
+\frac{h}{2}\cos\tilde\theta =0\nonumber\\
\end{eqnarray}
whose solution is
\begin{eqnarray}
\cos\tilde\theta_\pm=\frac{-\hbar\tilde\omega\pm\frac{\xi}{3}\sqrt{
h^2(1+\frac{\xi^2}{3})-3\hbar^2\tilde\omega^2}}
{\frac{h}{2}(1+\frac{\xi}{3})}\equiv 1-\Delta_\pm,
\end{eqnarray}
A similar procedure for the transitions from $hh-$ to $lh\pm$ yield
$\cos\tilde\theta_\pm=-1+\Delta_\mp$.
\begin{widetext}
Combining the contributions for each transition we then obtain
for $m_{lh}E_F/\mu-(h/6)m_{lh}/m_{hh}-h/2<\omega
<m_{lh}E_F/\mu+(h/6)m_{lh}/m_{hh}+h/2$
\begin{eqnarray}
A_{xy}(\omega)&=&
\int_{1-\Delta_+}^1d(\cos\theta) A_{xy}
(\omega,\cos\theta;hh+\rightarrow lh+)
+\int_{1-\Delta_-}^1d(\cos\theta) A_{xy}
(\omega,\cos\theta;hh+\rightarrow lh-)\nonumber\\
&&+\int_{-1}^{-1+\Delta_-}d(\cos\theta) A_{xy}
(\omega,\cos\theta;hh-\rightarrow lh+)
+\int_{-1}^{-1+\Delta_+}d(\cos\theta) A_{xy}
(\omega,\cos\theta;hh-\rightarrow lh-)
\nonumber\\
&=& -\frac{e^2\sqrt{2\mu\omega/\hbar}}{(2\pi)^2\hbar}
\int_{1-\Delta_+}^1d(\cos\theta)
\left(\frac{3}{4}\cos\theta\cos^2\theta' +\frac{h }{4\hbar \omega} [
-\frac{1}{2}\sin(2\theta)\sin(2\theta')
+2\cos2\theta\cos^2\theta'\right.\nonumber\\&&\left.\hskip 5 cm
+ \frac{\cos^2\theta\cos^2\theta'}{4\cos{2\theta'}} -\frac{3}{4}
\cos^2\theta\cos^2\theta' ]\right)\nonumber\\
&&-\frac{e^2\sqrt{2\mu\omega/\hbar}}{(2\pi)^2\hbar}
\int_{1-\Delta_-}^1d(\cos\theta)
\left(\frac{3}{4}\cos\theta\sin^2\theta' +\frac{h }{4\hbar \omega} [
+\frac{1}{2}\sin(2\theta)\sin(2\theta')
+2\cos2\theta\sin^2\theta'\right.\nonumber\\&&\left.\hskip 5 cm
- \frac{\cos^2\theta\sin^2\theta'}{4\cos{2\theta'}} -\frac{3}{4}
\cos^2\theta\sin^2\theta' 
]\right)\nonumber\\
&=& -\frac{e^2\sqrt{2\mu\omega/\hbar}}{(2\pi)^2\hbar}\left[
\left.\left(\frac{3}{8}u^2 +\frac{h }{4\hbar \omega} 
[ \frac{7}{6}u^3-2u ]\right)\right|_{1-\Delta_+}^1
\right.\nonumber %\\&&
\left.
+\left.\left(\frac{u}{8}\left(\sqrt{1-\frac{3}{4}u^2}
+\frac{3}{2}u\right)-\frac{\sqrt{3}}{12}
\arcsin(\sqrt{3}u/2)\right)\right|_{1-\Delta_-}^{1-\Delta_+}
\right.  \nonumber\\
&&\left.+\left.\frac{h }{4\hbar \omega} 
\left(-u+\frac{7u^3}{12}+\sqrt{1-\frac{3}{4}u^2}
\left(\frac{7}{27}+\frac{13u^2}{18}\right)
\right)\right|_{1-\Delta_-}^{1-\Delta_+} \right]\nonumber.
\end{eqnarray}
\end{widetext}
A similar procedure for the upper range 
$m_{hh}E_F/mu-h/6-(h/2)m_{hh}/m_{lh}$$<$$\omega$$<$$m_{hh}E_F/mu+h/6+(h/2)m_{hh}/m_{lh}$
yields $A_{xy}(\omega)$ given in Eq. \ref{Aomhigh}.
For any other value of $\omega$ $A_{xy}(\omega)=0$.

%\bibliography{dms}

\begin{thebibliography}{36}
\bibitem[{not({\natexlab{b}})}]{note3}
\bibinfo{note}{Energetic readers will discover that these calculations start
  with expressions as long as a west Texas dirt road but reach a relatively
  concise conclusion.}
\end{thebibliography}

%%%%the references below have been generated by bibtex:
\end{document}